\documentclass[12pt,preprint]{aastex}

\shorttitle{Gamma-Ray Burst $E_{peak}$ Distribution}
\shortauthors{Schaefer}

\begin{document}

\title{Explaining the Gamma-Ray Burst $E_{peak}$ Distribution}

\author{Bradley E. Schaefer}
\affil{Department of Astronomy, University of Texas,
    Austin, TX 78712}

\begin{abstract}

The characteristic photon energy for Gamma Ray Bursts, $E_{peak}$, has a
remarkably narrow distribution for bursts of similar peak flux, with
values between 150 and 600 keV for most faint bursts.  This result is
surprising within the framework of internal shock models, since spectral
shifts associated with the jet's blue shift (by a Lorentz factor of
$\Gamma$) and the cosmological red shift (by a factor of $1+z$) should
cause substantial smearing in the distribution of the spectral peak in the
jet's co-moving frame, $E_{rest}$.  For the general case where the
luminosity ($L$) varies as $\Gamma^N$ and $E_{rest}$ varies as $\Gamma^M$,
then the observed $E_{peak}$ will vary as $L^{(M+1)/N}(1+z)^{-1}$.  For
two independent set of 20 and 84 bursts, $E_{peak}(1+z)$ varies as a power
law of the luminosity with an index of $(M+1)/N=0.36 \pm 0.03$.  With this
measured value, the above functional dependence of $E_{peak}$ on $L$ and
$z$ results in $E_{peak}$ being roughly constant for bursts of similar
peak flux, $P_{256}$.  Thus, the kinematic smearing will be small, hence
allowing the $E_{peak}$ distribution to be narrow.  This model also
predicts that bright bursts will have high $E_{peak}$ values because they
all have some combination of high luminosity (and hence a large blue shift
$\Gamma$) and a nearby distance (and hence a small cosmological red
shift).  Quantitatively, $E_{peak}$ should vary roughly as
$P_{256}^{~~~0.36}$, and this model prediction is strikingly confirmed
with BATSE data by Mallozzi et al.  A prediction of this model is that
GRBs at very high red shift ($z \sim 10$) should all appear with
$E_{peak}$ at $\approx 200~keV$.  A further prediction of this model is
that normal bursts with $P_{256}$ below the BATSE trigger threshold will
appear as x-ray flashes with $E_{peak} \approx 70~ keV$; just as is
reported by Kippen et al. and Heise et al.

\end{abstract}



\section{Introduction}

Gamma-Ray Bursts (GRBs) are indeed 'gamma-ray' bursts since their spectrum
is dominated by gamma radiation.  $E_{peak}$ is the characteristic photon
energy measured from a burst spectrum by fitting a smoothly broken power
law (Band et al. 1993).  \citet{mal95} have measured $E_{peak}$ values for
399 BATSE bursts.  They find that a histogram of $E_{peak}$ is roughly
log-normal in shape, with average values of $\sim 200~keV$.  They also
find that the width of this distribution is surprisingly narrow, with a
dispersion corresponding to a factor of $\sim 2$ about the average.  
\citet{bra98} have proven that the narrowness of this distribution is not
an artifact of BATSE detector properties or of the analysis procedures,
and in particular that the sharp falloff in frequency down to at least 50
keV is certain.  Further, deep searches with the Solar Maximum Mission 
Gamma Ray Spectrometer have proven that the population of high-$E_{peak}$ 
bursts is nearly zero \citep{has98}.

This is not to say that 'GeV bursts' and 'keV bursts' do not exist. For
example, GRB950425 is certainly a GeV burst \citep{sch98}.  Also,
\citet{kip01} and \citet{hei01} report x-ray transients that have similar 
light curves and durations as classical GRBs.  The existence of these 
x-ray transients might be interpreted as implying that the $E_{peak}$ 
distribution has a broad tail to low energies, despite the iron-clad data 
from BATSE.  The resolution lies in the realization that the x-ray 
transients all have peak fluxes (in the 50-300 keV band) that are 
far below the normal BATSE trigger threshold.  The result from 
\citet{mal95} is that the $E_{peak}$ distribution is narrow {\it for 
bursts in a small range of peak flux}.  Bursts with peak fluxes far below 
the BATSE threshold do not (and should not) enter into the histograms 
constructed from BATSE triggers.  If an $E_{peak}$ distribution is 
constructed for very low peak flux events (including the x-ray 
transients), then a simple extrapolation of the average $E_{peak}$ versus 
peak flux relation of Mallozzi suggests that such events will appear as 
ordinary bursts except with $E_{peak} \sim 70$ keV, just as observed.

So apparently there is something about GRBs (the $E_{peak}$ value) that
is fairly constant over all bursts.  This poses a double challenge to
theory, since there must be some physical mechanism which acts as a
thermostat and there must be some mechanism for balancing the kinematic
blue shifts (due to the jet's velocity towards Earth) and red shifts (due
to the cosmological expansion).  Based on the factor of $\sim 1000$
variation in burst luminosity, the Lorentz factor of the jet likely varies
by a large factor, while the cosmological red shift varies by over a
factor of 3; so we might expect the kinematic effects to produce a broad
$E_{peak}$ distribution even if all bursts had the same rest frame
temperature.

\citet{mal95} also found that the average $E_{peak}$ value varied
systematically with the apparent brightness of the burst.  The peak flux
used here, $P_{256}$, is measured over a 256 ms time scale within a
limited energy band from 50-300 keV \citep{fis94}.  The bright BATSE
bursts have $\langle E_{peak} \rangle$ around 339 keV while the faintest
BATSE bursts have $\langle E_{peak} \rangle$ around 175 keV.  The trivial
explanation of this effect as arising only from cosmological red shifts is
unreasonable since burst peak fluxes vary more due to the wide spread in
luminosity than due to distances.  So a second mystery is why $\langle
E_{peak} \rangle$ varies with $P_{256}$ as observed.

\section{The Narrow $E_{peak}$ Distribution}

Two kinematic effects will change the characteristic photon energy of the
emitting region in its own rest frame, $E_{rest}$, into the characteristic
photon energy as seen by Earth, $E_{peak}$.  The first is the blue shift
caused by the bulk velocity of the emitting region towards Earth.  This
will increase the detected characteristic energy by a factor of $\Gamma$.  
The second is the red shift caused by the usual cosmological expansion of
the Universe.  This will decrease the detected characteristic energy by a
factor of $(1+z)^{-1}$, where $z$ is the red shift.  So, 
\begin{equation}
	E_{peak} = E_{rest} \Gamma (1+z)^{-1}.			
\end{equation}
Here, the $E_{peak}$ distribution will be a convolution of the $E_{rest}$
distribution and the kinematic smearing effects from the $\Gamma
(1+z)^{-1}$ factor.

The $E_{rest}$ value can vary from pulse-to-pulse and burst-to-burst.  In
general, $E_{rest}$ might have a power law dependence on $\Gamma$.  So
\begin{equation}
	E_{rest} = E_0 \Gamma ^M,	 	
\end{equation}
where $M$ is the power law index.  The exponent $M$ might be zero if there
is some process which limits the temperature in the emitting region, or
$M$ might be unity if the kinetic energy of the jet is entirely converted
to thermal energy and then into gamma radiation.

Within generic internal shock scenarios, the burst isotropic luminosity,
$L$, varies as some power of $\Gamma$.  One power of $\Gamma$ comes from
the available energy in the shell.  Two more powers of $\Gamma$ come from
the blue shifting of photons in the usual $E^{-2}$ burst spectrum.  (The
50-300 keV power law index varies over a typical range of about unity in
width from burst-to-burst, with an average of close to -2 as shown in Fig.  
4 of Schaefer et al. (1998) and Fig. 46 of Schaefer et al. (1994).)  The
bunching of photons by the forward motion of the shell will not increase
the luminosity since the overall timing of sub-pulses (and thus the
distribution of photons within the peak) is determined by the central
engine which has no $\Gamma$ dependence.  (Alternatively, if a single
sub-pulse with observed duration longer than 256 ms dominates the peak
flux, then an additional $\Gamma^2$ dependence is due to the bunching of 
photons.)  Hence, the luminosity should vary as something like $\Gamma^3$.  
However, variations on this basic scenario are possible; for example the
peak luminosity might well be dominated by a relatively small emission
area (to account for the fact that most bursts have millisecond variations
\citep{wsf00}) and the resulting relativistic beaming could provide
another factor of $\Gamma^2$.  Emission efficiencies could well have some
dependency on $\Gamma$.  To allow for any reasonable scenario, let me 
adopt a general expression of
\begin{equation}
	L = L_0 \Gamma^N,
\end{equation}
where $N$ is the power law index that is expected to be $\sim 3$ or
perhaps $\sim 5$ and to vary somewhat from burst-to-burst.

	Equations 1-3 can be combined as
\begin{equation}
	E_{peak} = E_0 (L / L_0)^{(M+1)/N} (1+z)^{-1}.
\end{equation}
The kinematic part can be isolated as	
\begin{equation}
	E_{peak} \propto L^{(M+1)/N} (1+z)^{-1}.
\end{equation}
This representation is convenient since $L$ and $z$ can be measured for
many bursts, either by direct measures of their optical red shifts or
based on the lag and variability distance indicators \citep{sdb01}.  This
is also convenient since lines of constant $P_{256}$ are also functions of
$L$ and $z$, and the narrow $E_{peak}$ distributions of \citet{mal95} were
constructed for bursts in narrow bins of $P_{256}$.  Figure 1 displays
curves of constant $E_{peak}$ (for the case of $(M+1)/N=0.36$) and lines
of constant $P_{256}$.  The two $P_{256}$ lines are for values ($10.0$ and
$1.0~ph~s^{-1} ~ cm^{-2}$) which are the medians of Mallozzi's brightest
and dimmest bins (for $\Omega_M=0.3, \Omega_{\Lambda}=0.7, H_{\circ}=65
~km~s^{-1}~Mpc^{-1}$).  The two curves of constant $E_{peak}$ values (339
and 175 keV) were chosen to correspond to the measured values for
Mallozzi's brightest and dimmest bins (for $L_0$ and $E_0$ normalized so
that the lower curve is close to the faintest $P_{256}$ line).

	Over the range for which GRBs are seen (typically $0.4<z<4$), the
$E_{peak}$ curves are fairly close to the $P_{256}$ lines.  This is to say
that for a collection of bursts within a narrow $P_{256}$ bin, the
kinematic shifts do not vary much.  Over a wide range of $L$ or $z$, the
overall shifting of $E_{rest}$ is roughly constant.  This constancy is an
essential compensation between the red and blue shifts, in that bursts
with low blue shifts (and hence low luminosities) must necessarily be
close to Earth and have offsetting low red shifts; while the high blue
shift jets (with high $L$)  are distant with an offsetting high red shift.  
So the combined kinematic smearing varies little, thus allowing GRBs to
have a narrow $E_{peak}$ distribution despite having a wide range of red
shifts and luminosities.

	A narrow $E_{peak}$ distribution requires both that the kinematic
smearing be small and that the $E_{rest}$ distribution be at least as
narrow.  The explanation just given answers only part of the requirements.  
A physical mechanism to explain why the $E_{rest}$ distribution is so
narrow is still needed (such as in Kazanas, Georganopoulos, \& 
Mastichiadis 2002).

The size of the kinematic smearing depends on the exponent of $L$ in Eq.
5.  There will be additional smearing due to the finite width of the
$P_{256}$ bins.  With the model from Eq. 5, I have calculated the total
smearing across the bin for the faintest bursts ($0.95<P_{256}<1.3$) over
the normal range of red shifts ($0.4<z<4$).  For an acceptable value of
$(M+1)/N$, the smearing across this bin should be smaller than the
observed spread of $E_{peak}$.  The FWHM of the $E_{peak}$ distribution
for the faintest bursts stretches from roughly 150 keV to 600 keV, a
factor of 4 \citep{mal95}.  The kinematic smearing is less than a factor
of 4 for all values of $(M+1)/N$ less than 0.45.  The least smearing (at
19\%) occurs for $(M+1)/N=0.21$.

\section{How $E_{peak}$ Varies With $P_{256}$}

	\citet{nem94} discovered that bright bursts are harder than faint
bursts, while \citet{mal95} showed specifically how $\langle E_{peak}
\rangle$ varies with $P_{256}$ (see Fig. 2).  The general formalism from
the previous section can yield a prediction for how $\langle E_{peak}
\rangle$ varies with $P_{256}$.  The reason for this variation is a
consequence of Eq. 5.  From Fig. 1, we see that there is nearly a
one-to-one correspondence between $\langle E_{peak} \rangle$ and
$P_{256}$.  But the exact predicted function will depend on the
distribution of bursts within a $P_{256}$ bin.  If the median burst is
roughly at a constant red shift (specifically, at the strong inflection in
the burst number density at $z \approx 1$ \citep{sdb01}), then Eq. 5
becomes
\begin{equation}
	E_{peak} \propto P_{256}^{~~(M+1)/N},
\end{equation}
since $L \propto P_{256}$.  Alternatively, if the median burst is roughly
at a constant luminosity (specifically, at the break in the GRB luminosity
function around $2 \times 10^{52} erg~s^{-1} cm^{-2}$ \citep{sdb01}), then
Eq. 5 takes on a substantially more complex functional form that is
similar to Eq. 6.  The likely case is midway between these two extreme
alternatives, with the median burst in each $P_{256}$ bin along a line
roughly perpendicular to the line of constant $P_{256}$.  Here, there is
no simple expression although the $E_{peak}$ distribution is close to a
power law in $P_{256}$.

The $\langle E_{peak} \rangle$ data from \citet{mal95} is displayed in
Figure 2.  What value of $(M+1)/N$ best fits the data based on Eq. 6?  
The chi-square is minimized for $(M+1)/N$ equal to $0.40 \pm 0.14$.  
Here, the median burst in each $P_{256}$ bin is taken to lie on a line
perpendicular to the line of constant $P_{256}$ in the $L$ versus $z$
plot.  In the next section, I will show that $(M+1)/N$ is equal to $0.36
\pm 0.03$, and this is still an acceptable fit (see Fig. 2).

The point of this section is that Eq. 5 provides a simple and general
explanation for the observed variation of $\langle E_{peak} \rangle$ with
$P_{256}$ (Eq. 6).  In words, the brighter bursts will have some
combination of higher luminosities (with a higher $\Gamma$) and be closer
(with a smaller $1+z$) than fainter bursts, so that the jet's blue shift
and cosmological red shift will work together to give a higher $E_{peak}$
for the brighter bursts.

\section{The Values of $M$ and $N$}

The values of $M$ and $N$ can tell us about the physics of gamma ray
bursts.  Observations can place reliable constraints on both $M$ and $N$.  
Sections 2 and 3 have already provided limits, and two more observational
constraints will now be presented.

For the 112 BATSE bursts with known $L$ and $z$ (based on the measured lag
and variability) reported in \citet{sdb01}, the $E_{peak}$ value at the
time of peak flux is known for 84 events.  (The reason for why the 28
events do not have a measured $E_{peak}$ is not known due to the untimely
death of R. Mallozzi.)  Eq. 5 shows that $E_{peak}(1+z)$ should vary with
$L$ as a power law with index $(M+1)/N$.  As shown in Figure 3, the
logarithms of $E_{peak}(1+z)$ and $L$ are indeed accurately related
linearly.  A chi-square fit gives $(M+1)/N=0.36 \pm 0.03$.

This same method can be applied to 20 bursts which have the luminosities
based on optically measured red shifts.  Figure 3 shows these bursts, with
the data taken from the literature (primarily \citet{sch02} and
\citet{ama02}).  This sample also shows a power law dependence, with a
highly significant correlation ($r=0.90$, for a chance probability of $3
\times 10^{-8}$).  The slope is $(M+1)/N$, which is fitted to be $0.38 \pm
0.11$.

The variability/luminosity relation \citep{frr00} can be used to constrain
$N$.  Variability ($V$) is a measure of how 'spikey' a burst light curve
is, which empirically varies as the inverse of the rise times in the light
curves.  The rise times will be limited by the geometric delays which
scale as $\Gamma^{-2}$, so $V \propto \Gamma^2$.  \citet{sch02} has shown
from 9 bursts with measured red shifts that $L \propto V^{1.57 \pm 0.17}$.  
Thus, we have $L \propto \Gamma^{3.14 \pm 0.34}$, which is to say $N=3.14
\pm 0.34$.

I now have four observational restrictions and two theoretical
restrictions on $N$ and $M$.  (1) The narrowness of the observed
$E_{peak}$ distribution requires $(M+1)/N<0.45$.  (2) The variation of
$\langle E_{peak} \rangle$ with $P_{256}$ is easily explained if
$(M+1)/N=0.40 \pm 0.14$.  (3)  The variation of $E_{peak}(1+z)$ with L
for two independent set of bursts shows that $(M+1)/N$ equals $0.36 \pm
0.03$ and $0.38 \pm 0.12$. (4) The variability/luminosity relation gives
$N=3.14 \pm 0.34$.  (5)  Ordinary relativistic effects for any internal
shock model implies that $N \ga 3$.  (6)  The conversion of the jet
energy to thermal energy suggests that $0 \ge M \ge 1$.  Constraints 1-3
reduce to $(M+1)/N=0.36 \pm 0.03$.  Constraints 4 and 5 give $N=3.14 \pm
0.34$.  These two results then imply that $M=0.13 \pm 0.15$, which is
consistent with constraint 6.  In round numbers, $N=3$ and $M=0$.

\section{Implications}

What should very high red shift ($z \sim 10$) GRBs look like?  A first
guess \citep[e.g.,][]{blo01} might be that the cosmological red shift will
transform them to appear more like x-ray bursts with $E_{peak} \sim 20$
keV.  The analysis in this paper shows that this first guess is wrong
since the $z \sim 10$ bursts must have a very high luminosity and hence a
very high $\Gamma$ which will blue shift the $E_{peak}$ back to gamma-ray
energies.  In particular, from either equation 5 or Figure 1, we see that
the highest red shift bursts will all have the same $E_{peak}$ as the
nearby events of the same $P_{256}$.  Thus, any $z>5$ bursts that are in
the BATSE catalog will have $E_{peak} \sim 200 keV$.

\citet{kip01} have used BATSE real-time data to look at fast x-ray
transients discovered by the BeppoSAX satellite \citep{hei01}.  They find
that these events have similar light curves and durations as GRBs, which
is suggestive that the events are normal bursts.  Their peak flux is up to
3 times lower than the BATSE trigger threshold while their median
$E_{peak}$ value is 70 keV.  The natural suggestion was made that very
faint and very soft bursts might be at very high red shift.  However, the
model presented in this paper shows that {\it all} bursts below the BATSE
threshold will on average have a low $E_{peak}$, with no necessity of high
red shift.  (This is also realized from a simple extrapolation of the data
from Mallozzi et al.  as shown in Figure 2.)  That is, a low $P_{256}$ is
due to some combination of low luminosity (hence a low blue shift from the
jet) and large distance (hence a large cosmological red shift) which will
produce systematically low $E_{peak}$ values.  While some of the fast
x-ray transients might be at high $z$, the steepness of the GRB luminosity
function implies that almost all are at moderate red shift.

It is disappointing that lines of constant $E_{peak}$ are closely parallel
to lines of constant $P_{256}$ in the $L$ versus $z$ plot.  If this had
not been true, then a simple measurement of $E_{peak}$ and $P_{256}$ would
define the bursts' position in the plot and we would hence know the burst
luminosity and red shift.

The value of $E_0$ is approximately constant for all bursts.  This
constancy is similar to recent results that bursts are 'standard candles'
from the lag/luminosity and variability/luminosity relations as well as
that the total energy of bursts is nearly a constant.  Thus it now seems
that many aspects of GRBs are constant, despite the chaotic appearance of
their light curves.  One implication of this constancy is that GRBs might
prove useful as tools for cosmology.

The lag/luminosity and variability/luminosity relations will be the
primary methods of getting luminosity distances.  Unfortunately, these
relations are not as tight as we might hope.  It is possible that
$E_{peak}$ might be used as a third parameter to improve the distance
indicators, much as the Cepheid luminosities are improved with a
period/color/luminosity relation.  Thus, lag/$E_{peak}$/luminosity and
variability/$E_{peak}$/luminosity relations might substantially improve
the accuracy of GRBs as standard candles.

\acknowledgments

I thank Robert Mallozzi and the BATSE Team for providing the $E_{peak}$
values for the 84 bursts.

\clearpage

\begin{figure}
\columnwidth=0.8\columnwidth
\plotone{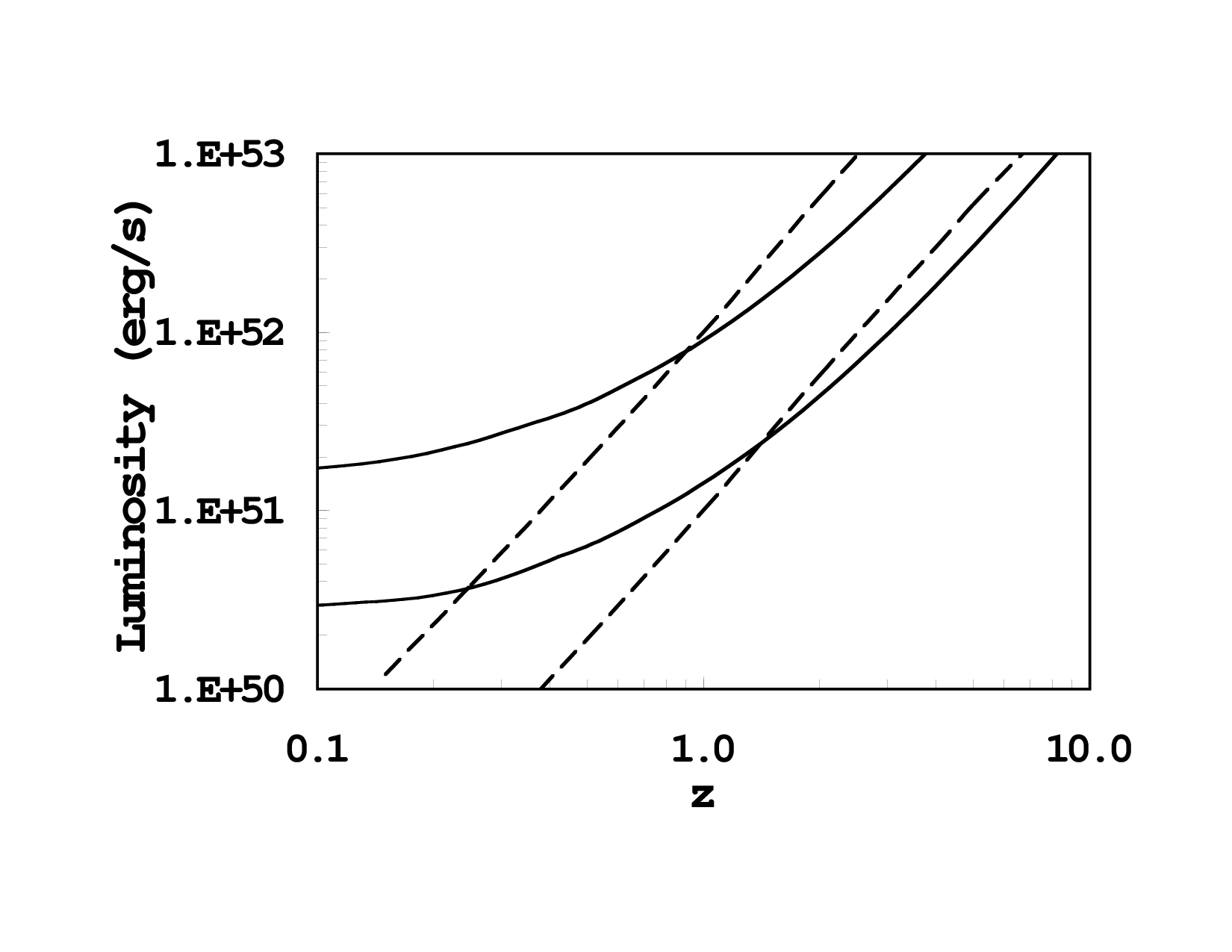}
\columnwidth=1.25\columnwidth
\vspace{-0.8in}
\caption{Lines of constant $E_{peak}$ and $P_{256}$.  In a plot of luminosity versus red 
shift, bursts of constant peak flux will fall on a nearly straight line.  Plotted are 
two dashed lines for  $P_{256}$ equal to $1.0$ and $10.0~photons~cm^{-2}~ s^{-1}$, which 
corresponds to the center of the faintest and brightest bins used by \citet{mal95}.  Also 
plotted are two curves of constant $E_{peak}$, as taken from Eq. 5 for $(M+1)/N=0.36$.  
The normalization factor has been chosen such that the lower curve with $E_{peak}=175~keV$ 
(corresponding to the faintest bursts) overlaps the $P_{256}$ bin for the faintest bursts.  
The upper curve is calculated for 
$E_{peak}$=339 keV, which corresponds to Mallozzi's brightest bursts.  The main point 
from this figure is that the lines and curves are roughly parallel over the range of 
observed bursts (roughly $0.4<z<4$), and this means that the blue shifts (from the 
jet's motion towards Earth) largely cancels out the red shifts (from the Universal 
expansion) for bursts of similar peak flux.  The curves and lines are not perfectly 
parallel which implies that kinematic effects will contribute to some of the observed 
scatter in the $E_{peak}$ distribution.  \label{fig1}}
\end{figure}

\clearpage 

\begin{figure}
\columnwidth=0.8\columnwidth
\plotone{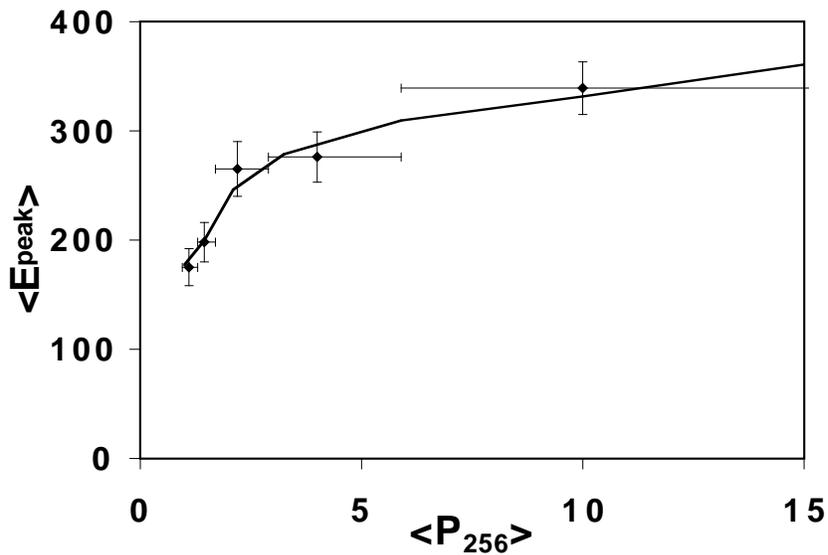}
\columnwidth=1.25\columnwidth
\caption{The model $E_{peak}$ as a function of $P_{256}$.  \citet{mal95} showed how the 
average $E_{peak}$ changes systematically with $P_{256}$, as shown by the plotted data 
in the figure.  From Fig. 1, we see that the model predicts a fairly close correspondence 
between $E_{peak}$ and $P_{256}$, with $E_{peak}$ increasing with $P_{256}$.  From Eq. 6, 
for the $(M+1)/N=0.36$ case with an arbitrary normalization, the model prediction is shown 
as a smooth curve.  The reason for the variation is that bright bursts are on average of 
higher luminosity (with  larger $\Gamma$ increasing the blue shift of the jet) and of 
nearer distance (with smaller z decreasing the cosmological red shift) than for fainter 
bursts.  There is a good match between model and observations.  \label{fig2}}
\end{figure}

\clearpage 

\begin{figure}
\columnwidth=0.8\columnwidth
\plotone{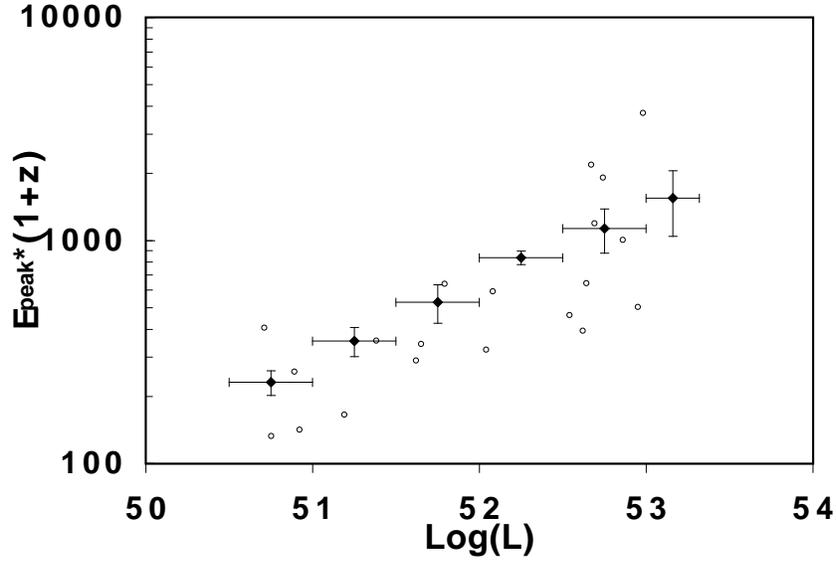}
\columnwidth=1.25\columnwidth
\caption{Direct fit of Eq. 5 to $E_{peak}$ data.  The quantity $E_{peak}(1+z)$ is related 
as a power law to the burst luminosity, and the slope will give the value for $(M+1)/N$.  
This is shown here for two independent data sets for which the luminosities are derived by 
two independent methods.  The first data set is for 20 bursts with  spectroscopically 
measured red shifts (the open circles).  The second is for 84 bursts (whose binned values 
are shown as filled diamonds) whose luminosity (and then red shift) were determined with 
the spectral lag and variability light curve parameters.  Both data sets show a highly 
significant and similar power law relation.  The existence of this relation demonstrates 
the validity of equation 5.  The explanation for this empirical relation  is that the high 
luminosity bursts have a high $\Gamma$ for their jet (which is why they are of high 
luminosity) which then blue shifts the $E_{rest}$ value to a high observed $E_{peak}$ 
value.  The fitted values for $(M+1)/N$ are $0.36 \pm 0.03$ and $0.38 \pm 0.11$ 
respectively.  \label{fig3}}
\end{figure}


\begin{thebibliography}{}
\bibitem[Amati et al. (2002)]{ama02} Amati, L. et al. 2002, \aap, 390, 81
\bibitem[Band et al.(1993)]{ban93} Band, D. L. et al. 1993, \apj, 428, 21
\bibitem[Bloom et al. (2001)]{blo01} Bloom, J. S., Frail, D. A., \& Sari, R. 2001, \aj, 121, 2879
\bibitem[Brainard et al. (1998)]{bra98} Brainard, J. J. et al. 1998, 
     20th Texas Symposium, Paris (astro-ph/09904039)
\bibitem[Fenimore \& Ramirez-Ruiz (2000)]{frr00} Fenimore, E. E.,
    \& Ramirez-Ruiz, E.  2000, \apj, submitted, (astro-ph/0004176)
\bibitem[Fishman et al. (1994)]{fis94} Fishman, G. J. et al. 1994, \apjs, 92, 229
\bibitem[Harris \& Share(1998)]{has98} Harris, M. J. \& Share, G. H. 1998, 
\apj, 494, 724
\bibitem[Heise et al. (2001)]{hei01} Heise, J., in't Zand, J., Kippen, R. M., \& Woods, P. M. 
2001, in Gamma-Ray Bursts in the Afterglow Era, eds E. Costa, F. Frontera, and J. Horth 
(Berlin; Springer), p. 16 (astro-ph/0111246)
\bibitem[Kazanas, Georganopoulos, \& Mastichiadis(2002)]{kgm02}
Kazanas, D., Georganopoulos, M., \& Mastichiadis, A. 2002, \apj,578, L15
\bibitem[Kippen et al. (2001)]{kip01} Kippen, R. M., Woods, P. M., Heise, J., int Zand, J., 
Preece, R. D., \& Briggs, M. S. 2001,  in Gamma-Ray Bursts in the Afterglow Era, eds E. Costa, 
F. Frontera, and J. Horth (Berlin; Springer), p. 22 (astro-ph/0102277)
\bibitem[Mallozzi et al.(1995)]{mal95} Mallozzi, R. S.,
    et al.  1995, \apj, 454, 597
\bibitem[Nemiroff et al. (1994)]{nem94} Nemiroff, R. J. et al. 1994, \apj, 435, L133
\bibitem[Schaefer (2002)]{sch02} Schaefer, B. E.
     2002, \apj, submitted
\bibitem[Schaefer, Deng, \& Band(2001)]{sdb01} Schaefer, B. E.,
    Deng, M., \& Band, D. L.  2001, \apj, 563, L123
\bibitem[Schaefer et al. (1998)]{sch98} Schaefer, B. E., J. Hobbeheydar, Bromm, V., 
\& Fenimore, E. 1998, in Gamma Ray Bursts, eds. C. A. Meegan, R. Preece, and T. Koshut 
(New York, AIP), p. 379
\bibitem[Schaefer et al.(1994)]{sch94} Schaefer, B. E. et al. 1994, 
ApJSupp, 92, 285
\bibitem[Schaefer et al.(1998)]{sss98} Schaefer, B. E. et al. 1998, \apj, 
492, 696
\bibitem[Walker, Schaefer, \& Fenimore (2000)]{wsf00} Walker, K. C., Schaefer, B. E., 
\& Fenimore, E. E. 2000, \apj, 537, 264
\end{thebibliography}
\end{document}